\documentclass[authoryear,5p,twocolumn]{elsarticle}
\usepackage{epsfig}
\usepackage{color}
\usepackage{amssymb}
\usepackage{amsmath}
\usepackage{amsthm}
\usepackage{multirow}
\usepackage{twoopt}
\usepackage{graphicx}
\definecolor{slateblue}{rgb}{0.2,0.2,0.6}

\usepackage[pdftex,breaklinks=true,colorlinks=true,linkcolor=slateblue,citecolor=slateblue,urlcolor=slateblue,allcolors=slateblue,pdfauthor={Tomassetti},pdftitle={CRASR19}]{hyperref}
\usepackage{xspace}

\newcommand{\ApJ}{Astrophys. J.}
\newcommand{\AeA}{Astron. \& Astrophys.}
\newcommand{\PRL}{Phys. Rev. Lett.}
\newcommand{\PRD}{Phys. Rev. D}
\newcommand{\PRC}{Phys. Rev. C}
\newcommand{\ASR}{Adv. Space Res.}

\newcommand{\AMS}{\textsf{AMS}\xspace}
\newcommand{\etal}{et al.}
\newcommand{\eg}{\textit{e.g.}} 
\newcommand{\ie}{\textit{i.e.}} 

\newcommand{\R}{\ensuremath{\mathcal{R}}}
\newcommand{\M}{\ensuremath{\mathcal{M}}}

\newcommand{\p}{\textsf{p}\xspace}
\newcommand{\Hyd}{\textsf{H}\xspace}
\newcommand{\He}{\textsf{He}\xspace}
\newcommand{\pHe}{\textsf{p/He}\xspace}

\newcommand{\C}{\textsf{C}\xspace}

\newcommand{\Oxy}{\textsf{O}\xspace}

\newcommand{\BC}{\textsf{B}/\textsf{C}\xspace}

\def\Journal#1#2#3#4{{#1}\,{#2}, #3;}  
\usepackage[utf8]{inputenc}
\usepackage{verbatim}
\newcommand*{\doi}[1]{\href{http://dx.doi.org/#1}{doi: #1}}

\makeatletter
\journal{Advances in Space Research} 

\begin{document}
\begin{frontmatter} 
\title{Numerical modeling of cosmic-ray transport in the heliosphere\\ and interpretation of the proton-to-helium ratio in Solar Cycle 24} 
\author[pgall]{Nicola Tomassetti}
\ead{nicola.tomassetti@pg.infn.it}
\author[lip]{Fernando Bar\~{a}o}
\author[pgall]{Bruna Bertucci}
\author[pgall]{Emanuele Fiandrini}
\author[lip]{Miguel Orcinha}
\address[pgall]{Universit\`a degli Studi di Perugia \& INFN - Sezione di Perugia, I-06100 Perugia, Italy} 
\address[lip]{Laborat{\'o}rio de Instrumenta\c{c}{\~a}o e F{\'i}sica Experimental de Part{\'i}culas, P-1000 Lisboa, Portugal}

\begin{abstract}
  Thanks to space-borne experiments of cosmic-ray (CR) detection, such as the \AMS and PAMELA missions in
  low-Earth orbit, or the Voyager-1 spacecraft in the interstellar space, a large collection of
  multi-channel and time-resolved CR data has become available.
  Recently, the \AMS experiment has released new precision data, on the proton and helium fluxes in CRs,
  measured on monthly basis during its first six years of mission.
  The \AMS data reveal a remarkable long-term behavior in the temporal evolution of the proton-to-helium ratio at 
  rigidity $\R\equiv{p/Z}\lesssim\,$3\,GV.
  As we have argued in a recent work, such a behavior  may reflect the transport properties of low-rigidity CRs in the inteplanetary space.
  In particular, it can be caused by mass/charge dependence of the CR diffusion coefficient.
  In this paper, we present our developments in the numerical modeling of CR transport in the Milky Way and in the heliosphere. 
  Within our model, and with the help of approximated analytical solutions,
  we describe in details the relations between the properties of CR diffusion  and the time-dependent evolution of the proton-to-helium ratio.
\end{abstract}

\begin{keyword}
cosmic rays \sep ISM \sep heliosphere \sep interplanetary space \sep solar modulation
\end{keyword}
\end{frontmatter}
\parindent=0.5 cm

\section{Introduction}      
\label{Sec::Introduction}   
%
The vast majority of cosmic rays (CRs) observed near-Earth originate well outside the solar system,
from spectacular astrophysical phenomena such as supernova remnants or stellar winds.
When entering the heliosphere, these particles are
decelerated by the solar wind and deflected by the interplanetary magnetic fields. 
This makes the energy spectra of CRs in the inner heliosphere significantly different from those in the interstellar space.
Moreover, the effect is known to change periodically, with the changing
number of sunspots appearing on the solar corona. 

The observed change of the Galactic CR flux over the variable Solar Cycle is known as \emph{solar modulation} of CRs in the heliosphere.
Along with its implications in solar or plasma astrophysics, solar modulation is an important phenomenon for Galactic CR physics studies,
as it limits our ability to identify the CR sources or to search for dark matter annihilation signals \citep{Grenier2015}.
Moreover, predicting the variations of the CR flux in the heliosphere
represents a significant challenge for interplanetary space missions \citep{Kudela2000,Tomassetti2017TimeLag}.

The solar modulation effect can be studied using
various types of CR data
such as counting rates from neutron monitors or direct measurements of CR fluxes \citep{Usoskin2005,Usoskin2011,Bindi2017,Wiedenbeck2013}. 
In the past decades, precious information have been gained thanks to space missions CRIS/ACE \citep{Wiedenbeck2009}, IMP-7/8 \citep{GarciaMunoz1997},
\emph{Ulysses} \citep{Heber2009},  and Voyager-1 \citep{Cummings2016}.
The recent data from EPHIN/SOHO \citep{Kuhl2016}, PAMELA \citep{Adriani2013,Martucci2018}, and \AMS \citep{Aguilar2018PHe,Aguilar2018Leptons}
are giving a deep characterization of effect in the current Solar Cycle 24 \citep{Bindi2017}.
The solar modulation effect is known to be caused by a combination of basic processes
such as magnetic diffusion, drift motion, convection, and adiabatic energy losses \citep{Parker1965,Krymsky1964,Fisk1971,Jokipii1967}. 
All these processes are linked to the nature of the heliospheric medium and on the dynamics of charged particles in magnetic fields.
The diffusive propagation arises from the erratic motion of CRs through magnetic turbulence, \ie,
charged particle scattering off the small-scale irregularities of the magnetic fields.
In the heliosphere, these irregularities are frozen in the expanding solar wind, and continuously transported radially 
towards the outer heliosphere. This gives rise to a global convective motion, and adiabatic energy losses associated with it.
On the other hand, the large-scale global structure of the heliospheric magnetic field, and in particular the
so-called heliospheric current sheet, give rise to charge-sign dependent effects in CR transport such as
drift.
All these processes are well correlated with the 11-year quasiperiodical cycle of solar activity (\emph{tout court} Solar Cycle)
and, for charge-sign dependencies, with with the 22-year cycle of magnetic polarity \citep{Potgieter2013,Potgieter2014b}.
At microscopic level, all these processes are governed by the rigidity (momentum/charge ratio) $\R=p/Z$, 
\ie, the quantity that best describes dynamics of particle (gyro)motion in magnetic fields.
In principle, one can connect the spectrum of heliospheric turbulence
with the rigidity dependence of the diffusion mean free path $\lambda$ of CR
particles \citep{TeufelSchlickeiser2002}. 

In the recent measurements released by the \AMS experiment  \citep{Aguilar2018PHe},
the temporal dependence of the hadronic fluxes
shows new details of CR modulation that can be used to test CR transport. 
More precisely, \AMS has measured the temporal variation of the proton and helium fluxes
in CRs, between 2011 and 2017, with a 27-day time resolution and a precision level of 1\,\%.
When measured at the same value of rigidity \R, the CR fluxes of protons and helium
are found to have very similar fine-structures in terms of time and relative amplitude.
However, on yearly time scales, little differences have been observed between the two species. 
Such a difference between the proton and helium fluxes can be observed in the 
temporal evolution of their ratio, \pHe, evaluated at rigidity $R\lesssim$\,3\,GV.
From the \AMS data, the \pHe{} ratio shows a remarkable decrease between March 2015 and May 2017,
corresponding to the \emph{descending phase} of solar activity,
after the 2014 solar maximum and toward the 2019 solar minimum.
This period is also called \emph{recovery phase} for the Galactic CRs,
because during this phase the absolute fluxes of proton and helium are seen to increase steadily.
Possible explanations for the long-term \pHe{} feature involve
mass/charge dependent-effects in CR transport at low-rigidity,
either differences in the local interstellar spectra (LIS)
of CR proton and helium. In both cases, the \pHe{} ratio can be also
influenced by the $^{3}$\He--$^{4}$\He{} isotopic
mixing \citep{Aguilar2018PHe,Jokipii1967,Gloeckler1967,Biswas1967,Herbst2017,Gieseler2017}.
In a recent paper \citep{Tomassetti2018PHeVSTime}, we have show that the
long-term \pHe{} behavior observed by \AMS represents a
remarkable signature of the rigidity dependence of CR diffusion in the heliosphere
that is described by a diffusion coefficient of the type $K(\R)\propto\beta(\R)\lambda(\R)$.
As we have argued, the velocity factor $\beta$ gives rise to little mass/charge dependence of
CR diffusion in the low-rigidity limit, \ie, $\R\lesssim\,3$\,GV.
Similar results were also reported by other authors \citep{Corti2019,Boschini2019,Luo2019}.
In all these works, albeit based on different modeling, 
the conclusions are driven by the availability of new important measurements:
the Voyager-1 data collected outside the heliosphere,
and the \AMS{} multi-channel and monthly-resolved data, collected locally
and over a significant fraction of the Solar Cycle.
The improved modeling is presented in details in the rest of this paper.
It consists in an accurate calculation of the proton and helium LIS's (beyond the heliosphere),
along with their corresponding modulated fluxes near-Earth (in the inner heliosphere).
This paper is organized as follows.
In Sect.\,\ref{Sec::Observations}, we recapitulate all the data we used in our analysis.
In Sect.\,\ref{Sec::Propagation}, we present the calculations of the LIS, \ie, our improved modeling in Galactic CR propagation in the ISM.
In Sect.\,\ref{Sec::SolarModulation}, we describe in details our numerical model of solar modulation and our data-driven approach that we have followed in order to test CR diffusion.
In Sect.\,\ref{Sec::Results} we present the results and discuss the interpretation of the data using the numerical model and approximated calculations.
We also discuss the main limitations of our modeling approaches. Conclusions are drawn in Sect.\,\ref{Sec::Conclusions}.

\section{Observations}     
\label{Sec::Observations}  

The model developments presented in this paper are motivated by the recent release of a large
sample of CR data from the \AMS collaboration.
The data published by the \AMS collaboration include lepton, proton, antiparticle spectra,
primary nuclei, secondary to primary ratios and, more recently, flux time dependencies.
All these data are available for download at the Cosmic-Ray Database of the
\emph{Space Science Data Center} (SSDC) at \emph{Italian Space Agency} \citep{DiFelice2017}.
In particular, the collaboration reported time-resolved measurements of CR proton and helium fluxes
based on a total sample of 846 and 112 billion collected events, respectively  \citep{Aguilar2018PHe,Aguilar2018Leptons}.
Both proton and helium fluxes have been measured for the 79 Bartel's rotations (BR),
\ie, 27-day time intervals, ranging from May 2011 to May 2017. 
All proton fluxes range from 1 GV to 60 GV of rigidity. The helium fluxes range from 1.9 GV to 60 GV.
This sample represents a simultaneous measurement of an uninterrupted time-series of 
CR proton and helium fluxes over an extended period of time.
Overall, the variations of the proton and helium fluxes show similar fine structures, on monthly time scales.
It was also noted that these structures are well correlated with the monthly number of sunspot \citep{Corti2019},
(\ie, with the Sun's magnetic activity) and that their amplitudes decrease with increasing rigidity. 
With the precision of the \AMS measurements, time-variations in the individual \p-\He{} fluxes are appreciated up to nearly 40 GV of rigidity.
Most of these variations are canceled out in the flux ratio \pHe{} of proton and helium at the same value of rigidity,
The \pHe{} ratio is observed to be constant at rigidity above 3 GV.
Below 3 GV, however, the ratio show an interesting long-term time dependence. 
In particular, it decreases steadily in the period between Mar 2015 and May 2017, 
in the descending phase of the Sun's activity and during which the absolute flux intensity increases rapidly.

To interpret these measurements within a data-driven approach, other sets of data are needed such as combined data from \AMS and Voyager-1.
We used these data to constrain the interstellar CR spectra for proton and helium, and to estimate their uncertainties.
From the Voyager-1 experiment, we used the available interstellar data (\ie, collected after August 2012) down to 100 MeV/n energies.
In particular, hydrogen flux data are available in the energy range 140-320 MeV.
Data on CR helium and heavy ions data are available at 110-600 MeV/n \citep{Cummings2016}.
These data have been collected by Voyager-1 in the period 2012-2016, \ie, while the spacecraft were moving in the interstellar space.
Thus they provide direct and valuable constraints to CR propagation models.
From \AMS, we used data on the primary CR fluxes \p-\He-\C-\Oxy{} at $\R>$\,60\,GV \citep{Aguilar2015Proton,Aguilar2015Helium},
and secondary-to-primary \BC{} ratio at $\R>$\,4\,GV \citep{Aguilar2018LiBeB}.
The minimal rigidity values are chosen to ensure that the data are representative of the interstellar CR flux
(\ie, solar modulation free).
We note that between the Voyager-1 flux data (below $\sim$\,1\,GV) and the \AMS data (above $\sim$60\,GV), there
is a large energy gap where no information is directly available on the LIS flux.
Recovering the LIS fluxes inside the gap is possible, however, by means of a robust modeling of the Galactic CR propagation processes.
The \AMS{} data on primary fluxes and secondary-to-primary ratios are used for the determination of the model parameters,
and eventually for the determination of the LIS shape (Sect.\,\ref{Sec::Propagation}).

\section{Galactic cosmic-ray propagation}  
\label{Sec::Propagation}                   

\begin{figure}[!t]
\centering
\includegraphics[width=0.45\textwidth]{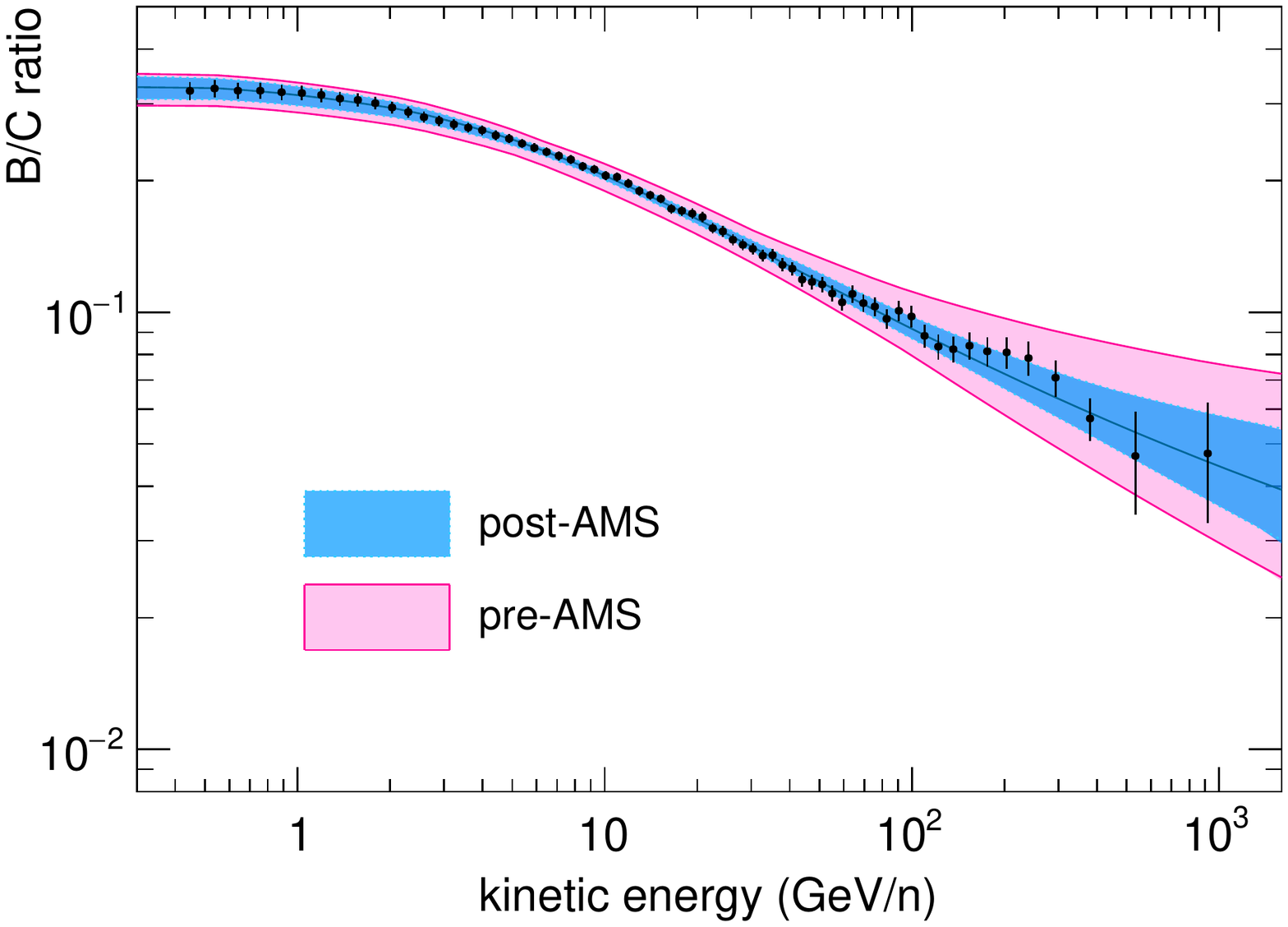}
\caption{ 
  Calculations of the \BC{} ratio and CR propagation uncertainties before and after
  the use of the new data from \AMS \citep{Aguilar2018LiBeB}.
}
\label{Fig::ccBCRatioVSEkn}
\end{figure}

The precise knowledge of the Galactic CR fluxes beyond the heliosphere is of crucial importance
for the study of solar modulation, and in particular the interpretation of the tiny features observed in the \pHe{} ratio. 
Traditionally, in solar modulation studies, the LIS fluxes constitute a mere physical input that
is assumed to be known, so they are often provided as parametric input functions.
The parametric approach was followed in all past studies
where, in fact, large discrepancies were noted among the different works \citep{Fisk1971,GleesonUrch1971,Moraal2013,Manuel2014,Herbst2017}.
A strong improvement was brought by the release of Voyager-1 data \citep{Corti2016,Corti2019,Herbst2017} 
To build a LIS parametric  model from the data, however, one has to deal with a large observational gap
between the Voyager-1 LIS data (available only at $\R\lesssim$\,1\,GV) 
and the \AMS data (\emph{modulation-free} only at $R\gtrsim$50\,GV) \citep{Corti2016}.
Hence, the use of a physically plausible shape for the LIS model is essential \citep{Boschini2017}.
For this reason, we opted for a full calculation approach based on a physically motivated model of CR propagation in the Milky Way.
The advantage of our approach is that it can provide a robust evaluation of the LIS uncertainties at all relevant energies.

To compute the Galactic CR fluxes of all particles, including proton and helium isotopes,
we made use of a spatial dependent model of diffusive CR propagation
in two halos \citep{Tomassetti2012Hardening,Tomassetti2015TwoHalo}.
In this model, the CR diffusion near the Galactic disk  is characterized
by a smaller rigidity dependence in comparison to CR diffusion in the halo away from the disk. 
The master equation describing CR transport in the Milky Way can be written as:
\begin{equation}\label{Eq::DiffusionTransport}
  \partial_{t} {N} = S + \vec{\nabla}\cdot (D\vec{\nabla}{N}) - {{N}}{\Gamma} + \partial_{E} (\dot{E} {N})  \,,
\end{equation}
where $N=N(E,r,z)$ is the particle number density as a function of energy and 2D spatial coordinates,
$\Gamma= \beta c n \sigma$ is the destruction rate for collisions off gas nuclei, with density $n$,
at velocity $\beta c$ and cross section $\sigma$. The source term $S$ is split into primary and secondary terms.
The former describes the CR acceleration spectrum from Galactic sources (\eg, supernova remnants) 
$S_{\rm pri}\propto(\R/{\rm GV})^{-\nu}$, where $\nu=$\,2.28$\pm$0.12 for protons and $\nu=$\,2.35$\pm$0.13 for all heavier nuclei.
The secondary production term for a given CR species is computed as $S_{\rm sec}= \sum_{\rm h} \Gamma_{h}^{\rm sp} N_{\rm j}$,
describing the fragmentation of all heavier $h$-nuclei occurring at rate $\Gamma_{h}^{\rm sp}$. 
The term $\dot{E}=\--\frac{dE}{dt}$ describes standard processes of ionization and Coulomb losses. 
While all sources and gas are located in the disk ($z=0$), the CR transport occurs in the halo up to a vertical extent $z=\pm\,L$.
The $L$ parameter sets the boundary conditions for the steady-state solution of Eq.\,\ref{Eq::DiffusionTransport}, $\partial_{t}{N}\equiv{0}$.
The CR transport is governed by a two-zone diffusion coefficient $D \equiv \beta D_{0}(\R/GV)^{\delta_{i/o}}$, where  $D_{0}/L=0.01\pm$0.002\,kpc/Myr.
The two diffusion regimes are described by the index ${\delta_{i/o}}$, that takes two values: $\delta_{i}=0.18\pm$0.05 
in the inner halo (near-disk region $|z|<\xi\,L$), and  $\delta_{o}=\delta_{i}+\Delta$
in the outer halo ($|z|>\xi\,L$), where $\xi=0.12\pm$0.03 and $\Delta=0.55\pm$0.11.
Once Eq.\,\ref{Eq::DiffusionTransport} is resolved for all CR species, giving $N=N(E,r,z)$, the local interstellar fluxes are 
computed as $J=\frac{\beta c}{4\pi}N_{\odot}$. Here the CR density is evaluated at the position of the Solar System $z_{\odot}\cong$0 and $r_{\odot}\cong$8.3\,kpc.
%
\begin{figure}[!t]
\includegraphics[width=0.47\textwidth]{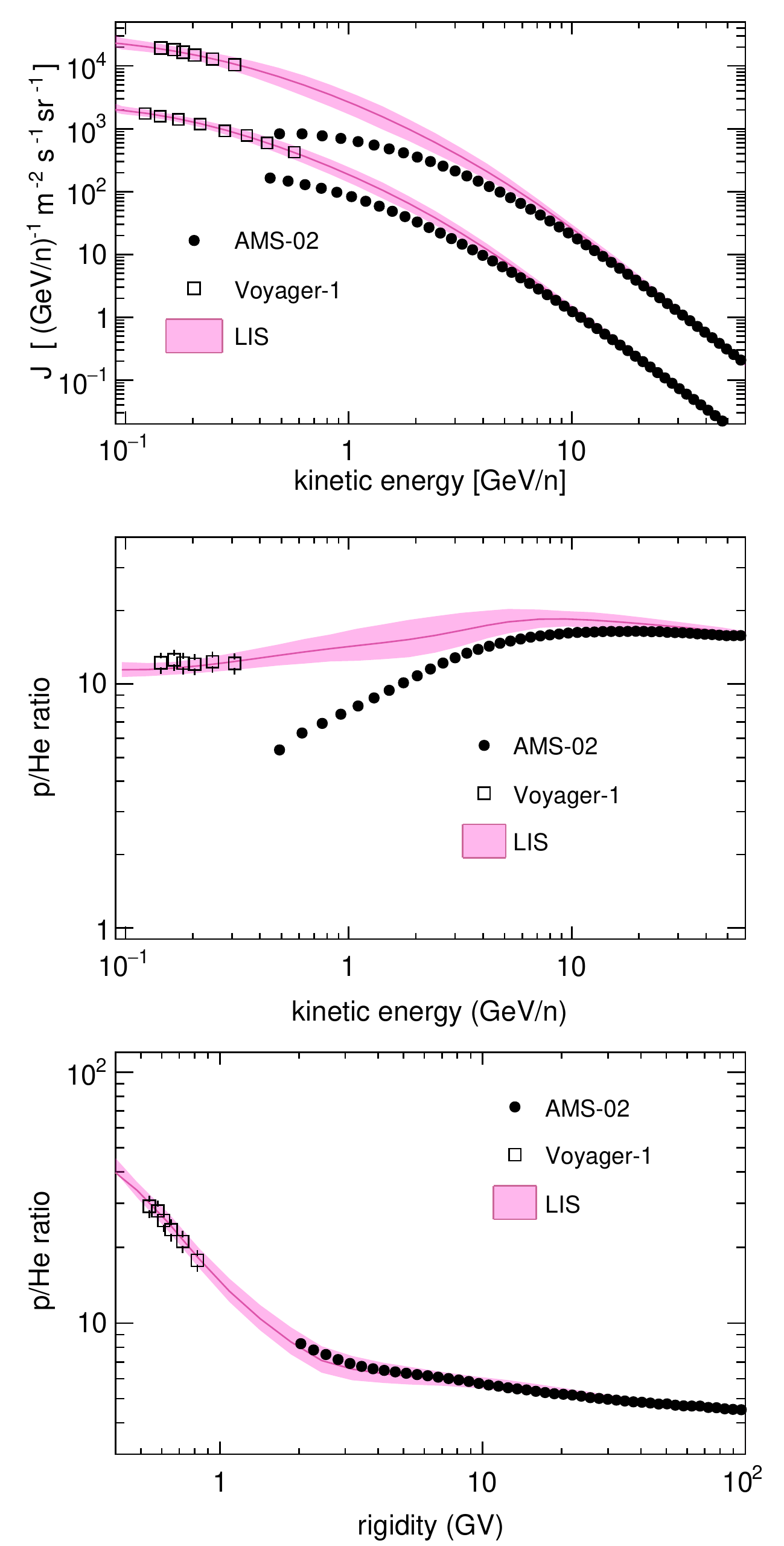}
\caption{ 
  (a) Measured and calculated proton and helium LIS's. The data from AMS \citep{Aguilar2015Proton,Aguilar2015Helium} and Voyager-1 \citep{Cummings2016};
  (b) Measured and calculated \pHe{} ratio as function of kinetic energy per nucleon;
  (c) Measured and calculated \pHe{} ratio as function of rigidity.
}
\label{Fig::ccLISProtonHelium}
\end{figure}
%
In all the above mentioned parameters, the best-fit values and their corresponding uncertainties have been
estimated by means of a Monte-Carlo Markov Chain sampling algorithm \citet{Feng2016}.
In addition to our benchmark setting, the effect of CR reacceleration in Alfv{\'e}nic waves was also tested. 
An advantage of reacceleration models is their explanation for the GeV peak in the \BC{} ratio. 
We found, however, that our model favors only modest values for the Alfv{\'e}nic speed, in agreement with the basic expectations
for the interstellar medium, from which $v_{A}\approx\,B/\sqrt{4\pi\rho}\lesssim$\,6\,km/s.
In many past studies based on homogeneous diffusion, however, large  
Alfv\'enic speeds were reported (up to $v_{A}\sim\,30-40$\,km\,s$^{-1}$) \citep{Trotta2011,Vladimirov2011}.
These models, however, require the introduction of
low-energy breaks in the injection spectra (at $E\sim$\,10\,GeV) in order to match the primary CR fluxes,
while additional high-energy breaks are needed to describe the observed spectral hardening at $E\sim$\,200\,GeV.
On the other hand, in our model, the full shape of primary CR spectra is interpreted naturally in terms of CR propagation in two diffusive regions.
Within this scenario, the GeV peak of the \BC{} ratio should be interpreted as a low-energy change of
CR diffusion, due \eg, to magnetic damping \citep{Ptuskin2006}. 
A critical discussion to CR reacceleration can be found in \citet{DruryStrong2017}. 

\begin{figure}[!t]
\centering
\includegraphics[width=0.45\textwidth]{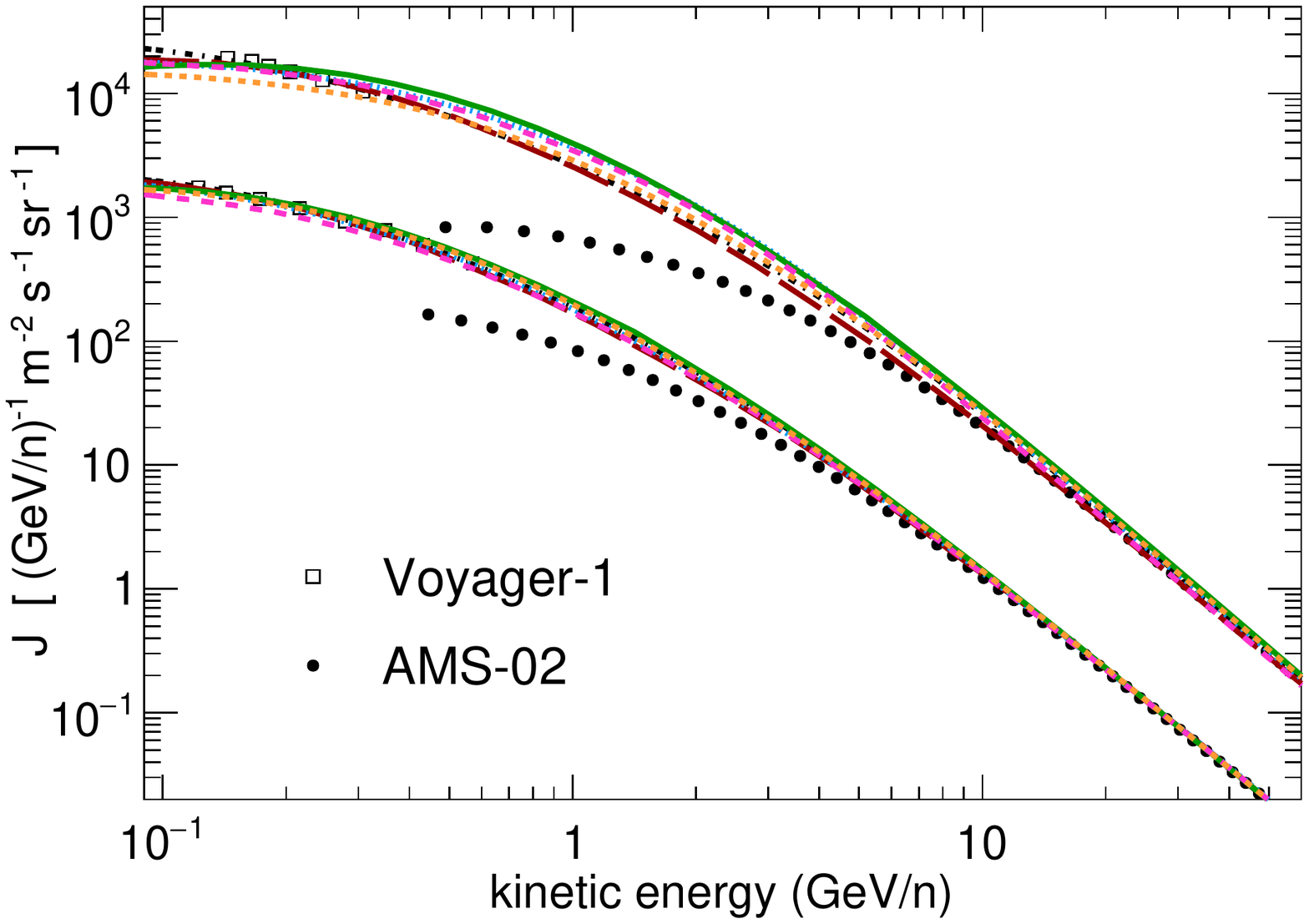}
\caption{ 
  Proton and helium
  data from \AMS and Voyager-1 \citep{Aguilar2015Proton,Aguilar2015Helium,Cummings2016},
  in comparison with LIS from various works:
  long-dashed red \citep{Tomassetti2017TimeLag},
  dashed orange \citep{Boschini2017}, 
  dot-dashed black \citep{Corti2016},
  dotted blue \citep{Tomassetti2017BCUnc},
  dashed pink \citep{Tomassetti2018PHeVSTime}, 
  and solid green line \citep{Tomassetti2015PHeAnomaly,Tomassetti2017Universality}. 
}
\label{Fig::ccLISComparison}
\end{figure}
%
Having modulation-free CR data is essential, in this phase of the work, to obtain unbiased LIS's. 
Thus, we have examined the presence and influence of potential biases caused by residual modulation.
In the LIS fitting procedure, the Voyager-1 data were used  in combination with the high-rigidity
\AMS data on primary CR fluxes \citep{Aguilar2015Proton,Aguilar2015Helium},
along with the newly released data on the \BC{} ratio \citep{Aguilar2018LiBeB}.
With the use of minimal rigidity thresholds (60 GV for the absolute fluxes and 4 GV for the \BC{} ratio \citep{Tomassetti2017BCUnc})
the effects of solar modulation are negligible at the $\lesssim$\,1\,\% level in the selected data.
Nonetheless, as in \citep{Feng2016}, we accounted for residual modulation
using a \emph{force-field} modulation potential $\phi$ as nuisance parameter.

The comparison between pre-\AMS \citep{Feng2016} and post-\AMS uncertainties is illustrated in Fig.\,\ref{Fig::ccBCRatioVSEkn}.
As shown in the figure, the new \AMS data on the \BC{} ratio pose tight constraints to the CR propagation parameters,
that are now comparable to the uncertainties arising from the fragmentation cross-sections  \citep{Tomassetti2015XS,Tomassetti2017BCUnc}. 
From the tight \BC-driven constraints, we obtained improved calculations of the secondary $^{2}$\Hyd{} and $^{3}$\He{} isotopes.
Their production is computed using an improved set of fragmentation cross-sections \citep{Tomassetti2012Iso,TomassettiFeng2017}.
In practice, the deuteron ($^{2}$\Hyd) contribution turned out to be fully negligible for all purpose, being
$^{2}$\Hyd/$^{1}$\Hyd{}$\lesssim$\,1\,\%.
On the other hand, the $^{3}$\He{} nuclei contribute by $\sim$\,15\,\%  to the total \He{} flux in the GV region.
Accounting for the $^{3}$\He{} is essential for a precise description of the CR helium data \citep{Biswas1967,Boschini2017}.
In agreement with \citet{Corti2019}, we found that it causes a moderate impact on the temporal dependence of the \pHe{} ratio.

In Fig.\,\ref{Fig::ccLISProtonHelium} the final proton and helium LIS's resulting from calculations are shown.
In the figure, the pink shaded bands represent the 1-$\sigma$ uncertainties estimated from the fit.
It can be seen that the highest uncertainties lie in the $\sim$\,1-10\,GeV energy region, where direct LIS data are not available.
This region is also sensitive to CR transport parameters such as $\delta_{i/o}$, $D_{0}/L$, or $v_{A}$. However,
proton and helium fluxes show very similar dependencies upon these parameters, 
so that the associated uncertainties are correlated.
As a result, the \pHe{} ratio  of Fig.\,\ref{Fig::ccLISProtonHelium}c show small uncertainties at all rigidities.
In particular, it is interesting to note that the \pHe{} ratio measured inside the heliosphere is very similar
to its interstellar value, when it is computed as a function of rigidity.
This explains why, before \AMS, no time variations were observed for the \pHe{} ratio \citep{Bindi2017}.

In Fig.\,\ref{Fig::ccLISComparison}, we plot some recent LIS derivations in comparison with the
data \citep{Tomassetti2018PHeVSTime,Tomassetti2017TimeLag,Corti2016, Tomassetti2015PHeAnomaly,Tomassetti2017BCUnc,Tomassetti2017Universality}.
These models are in agreement with our calculations within uncertainties, and they give consistent results.
The high degree of agreement among the different CR propagation models is mainly ascribed to the tight constraints of the recent data.
In particular, the Voyager-1  data have driven the CR LIS models to a new standard of precision in the low-energy region.
From the pre-Voyager-1 studies, it can be noted that the proposed LIS were highly discrepant
in the GeV and sub-GeV energy region \citep{Herbst2017}.

\begin{figure*}[!ht]
\centering
\includegraphics[width=0.90\textwidth]{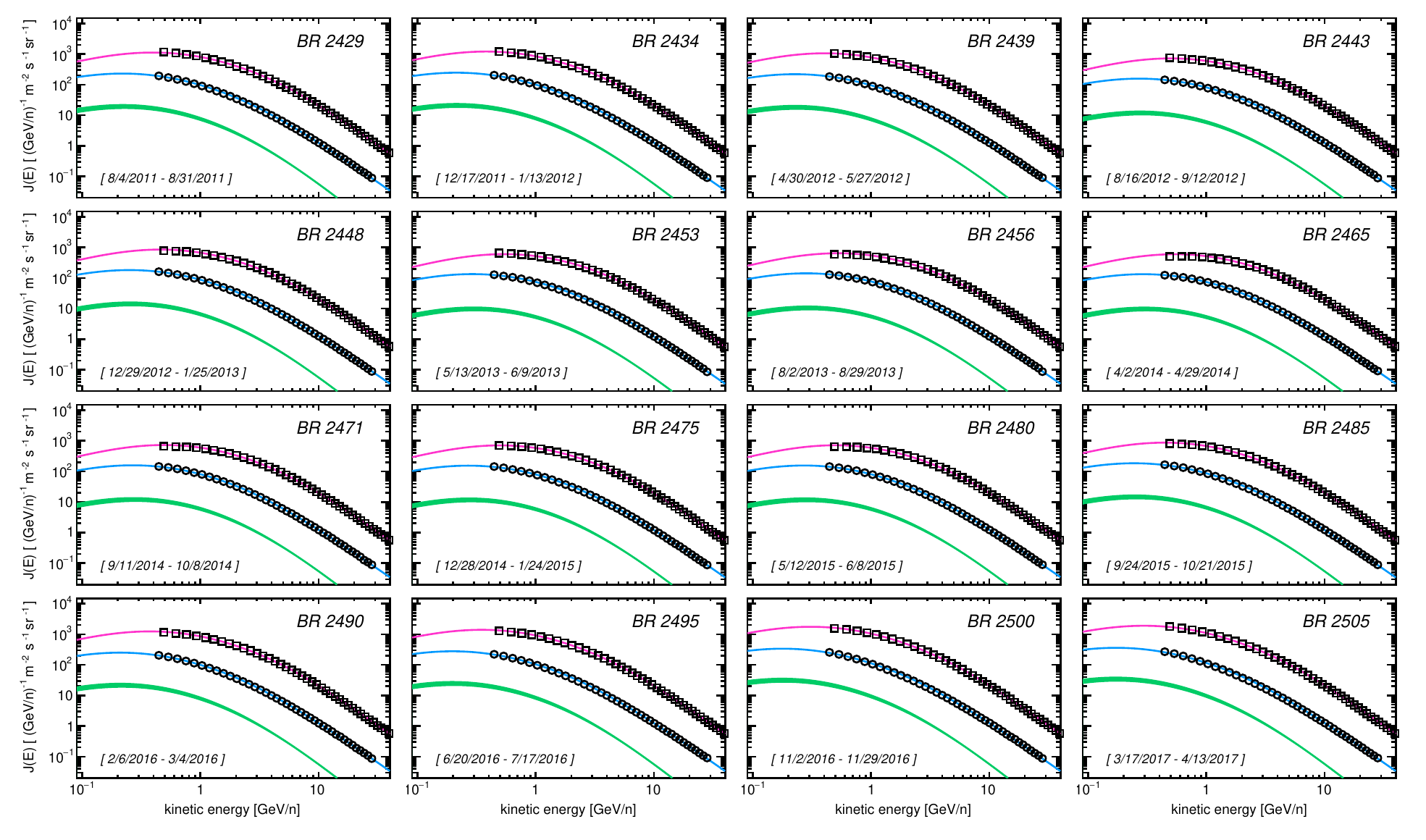}
\caption{ 
  Modulated energy spectra from 16 selected epochs.
  The calculations show the best-fit proton fluxes (pink), the predicted helium fluxes (blue), and the
  $^{3}$\He{} contribution (green shaded band, showing its corresponding uncertainties).
  The calculations are compared with the data from \AMS for the total fluxes of CR protons (squares) and helium (circles) \citep{Aguilar2018PHe}. 
} 
\label{Fig::ccEnergySpectra}
\end{figure*}

\section{Solar modulation calculations}    
\label{Sec::SolarModulation}               
%
The transport of cosmic rays (CRs) in the heliosphere is described by the Krymsky-Parker equation \citep{Krymsky1964,Parker1965}.
For a given CR particle type, the evolution of its phase-space density
$\psi=\psi({\bf r},\R,t)$ is governed by:
\begin{equation}\label{Eq::CRTransport}
\frac{\partial\psi}{\partial{t}}=\nabla\cdot\left(K\cdot\nabla\psi\right) - V\cdot\nabla\psi + \frac{1}{3}\nabla{V}\frac{\partial\psi}{\partial\rm{ln}\R} + Q
\end{equation}
where $\R$ is the particle rigidity.
The various terms represent convection with the solar wind of speed ${V}$,
spatial diffusion with tensor ${K}$, adiabatic momentum losses, and CR sources.          
Analytic solutions of this equation rely in simplifying approximations pertaining diffusion or losses.
In spherical symmetry, for a wind radially flowing at speed $V$ and isotropic diffusion tensor,
manageable numerical solutions can be found \citep{Fisk1971,GleesonUrch1971,Moraal2013}. 
The equation can be rewritten as:
\begin{equation}
\frac{\partial{\psi}}{\partial{t}}=\frac{1}{r^{2}}\frac{\partial}{\partial{r}}\left( r^{2}K\frac{\partial{\psi}}{\partial{r}}\right)
-V\frac{\partial{\psi}}{\partial{r}}
+\frac{1}{3r^{2}}\frac{\partial}{\partial{\R}} r^{2}V \frac{\partial{\psi}}{\partial{\rm{ln}\R}} +Q
\end{equation}
For steady-state conditions ($\partial\psi/\partial{t}=0$) and with zero source terms ($Q=0$), it becomes 
a parabolic equation of second order in space and first order in rigidity:
\begin{equation}
  K\frac{\partial{^{2}}\psi}{\partial{r^{2}}} +\left( \frac{\partial{K}}{\partial{r}}+\frac{2K}{r} -V \right)\frac{\partial{\psi}}{\partial{r}}+\left( \frac{2V}{3r}+ \frac{1}{3} \frac{\partial{V}}{\partial{r}} \right)\frac{\partial{\psi}}{\partial{\rm{ln}{\R}}}=0
\end{equation}
The boundary conditions can be set at $r=r_{\odot}$, the Sun's radius, and at $r=d$, the heliosphere's boundaries.
At $r=r_{\odot}$, the absence of any sources or sinks is assumed, so that ${\partial\psi}/{\partial{r}}|_{r_{\odot}}=0$ and the wind vanishes.
At the heliopause $r=d$, the condition is stated by requiring that the CR flux is equal to its local interstellar (LIS) value, so that $J_{T}|_{r=d}=\R^{2}\psi=J_{\rm LIS}$.
The initial condition results from the fact that the LIS is unchanged inside the heliosphere at
sufficiently high rigidity $\R_{\rm max}$:
$J(r,\R_{\rm max}) = J_{\rm LIS}(\R_{\rm max})$ for every $r$ between $r_{\odot}$ and $d$.
It should be noted that this condition accounts for the LIS at the boundaries ($\psi_{0}\equiv{Q}$) without the explicit specification of a source term $Q$ in the transport equation.
To obtain the steady-state solutions for $\psi(r,\R)$, we make use of the Crank-Nicolson implicit scheme.
In this method, the domain is divided in a rectangular grid with $r$ ranging from $r_{\odot}$ to $d$ and $\rm{ln}\R$
ranging from $\rm{ln}\R_{\rm min}$ to $\rm{ln}\R_{\rm max}$. 
The CR distribution function $\psi(r,\R)$ is represented as $\psi^{j}_{i}$ in
in the $N\times{M}$ space of coordinates $r_{i}=r_{\odot}+i\Delta_{r}$ and $\rm{ln}\R_{j}=\rm{ln}\R_{0}+j\Delta_{\rm{ln}\R}$, where $i=\,1\dots{N}$ and $j=\,1\dots{M}$.
The method consists of taking the average between the derivative evaluated at node $j$ and at node $j+1$, leading to discrete first and second order derivatives:
\begin{align}
&  \frac{\partial\psi}{\partial{r}} =  \frac{\psi^{j+1}_{i+1}-\psi^{j+1}_{i-1}}{4\Delta_{r}}+ \frac{\psi^{j}_{i+1}-\psi^{j}_{i-1}}{4\Delta_{r}} \\
&  \frac{\partial^{2}\psi}{\partial{r^{2}}} = \frac{\psi^{j+1}_{i+1}-2\psi^{j+1}_{i}+\psi^{j+1}_{i-1}}{2\Delta^{2}_{r}}+\frac{\psi^{j}_{i+1}-2\psi^{j}_{i}+\psi^{j}_{i-1}}{2\Delta^{2}_{r}}\\
&  \frac{\partial\psi}{\partial\rm{ln}\R} = \frac{ \psi_{i}^{j+1}-\psi_{i}^{j} }{\Delta_{\rm{ln}\R}}
\end{align}
After defining the quantities $\varphi_{1}=K$, $\varphi_{2}=-V+\frac{2K}{r}$,
and $\varphi_{3}=\frac{2}{3}\frac{V}{r}+\frac{1}{3}\frac{\partial{K}}{\partial{r}}$,
the equation becomes:
\begin{align}
&  \psi^{j}_{i-1}\left( \frac{\varphi_{1}}{2\Delta^{2}_{r}}-\frac{\varphi_{2}}{4\Delta_{r}}   \right) +
  \psi^{j}_{i}  \left(-\frac{\varphi_{1}}{\Delta^{2}_{r}}-\frac{\varphi_{3}}{\Delta_{\rm{ln}\R}} \right) +\\
&  \psi^{j}_{i+1}\left( \frac{\varphi_{1}}{2\Delta^{2}_{r}}+\frac{\varphi_{2}}{4\Delta_{r}}   \right) =
  \psi^{j+1}_{i-1}\left( -\frac{\varphi_{1}}{2\Delta^{2}_{r}}+\frac{\varphi_{2}}{4\Delta_{r}}   \right) +\\
&  \psi^{j+1}_{i}  \left( \frac{\varphi_{1}}{\Delta^{2}_{r}}-\frac{\varphi_{3}}{\Delta_{\rm{ln}\R}} \right) +  
  \psi^{j+1}_{i+1}\left(-\frac{\varphi_{1}}{2\Delta^{2}_{r}}-\frac{\varphi_{2}}{4\Delta_{r}}   \right)
\end{align}
By renaming the LHS coefficients as $\eta_{1}$, $\eta_{2}$, $\eta_{3}$, and the RHS one as $\eta_{4}$,
the system can be rewritten in a compact
form $A{\bf x^{j+1}}=B{\bf x^{j}}$, where $A$ and $B$ are tridiagonal matrices:
\[
\begin{bmatrix}
    \eta_{2} & \eta_{3} & 0 & \dots  & 0 \\[0.3pt]
    \eta_{1} & \eta_{2} & \eta_{3} & 0  & \dots \\[0.3pt]
    0 & \eta_{1} & \eta_{2} & \eta_{3}  & \dots \\[0.3pt]
    \vdots & \vdots & \vdots & \ddots & \vdots \\[0.3pt]
    0 & \dots & \dots & \eta_{2}  & \eta_{3}
\end{bmatrix}
\begin{bmatrix}
  \psi_{2}^{j}\\
  \psi_{3}^{j}\\
  :\\
  \vdots\\
  \psi_{N-1}^{j}\\  
\end{bmatrix}
=
\begin{bmatrix}
  \eta_{4}\\
  \eta_{4}\\
  :\\
  \vdots\\
  \eta_{4}-\eta_{3}\psi_{N}^{j}
\end{bmatrix}
\]
The partial differential equation is then transformed in a set of linear equation relating $\psi^{j}$ and $\psi^{j+1}$.
Every time the matrix is inverted, $\psi^{j}$ progresses one step in rigidity toward $\rm{ln}(\R_{\rm min})$.
The full solution is found in the whole domain by stepping down in rigidity and solving the tridiagonal system at each step.
In this work, we have set $N=$\,610 radius nodes, $M=$\,500 rigidity nodes, $\R_{\rm min}=$\,50\,MV and $\R_{\rm max}=$\,1\,TV for all CR particles.
The tridiagonal matrix equations are solved using an iterative method.
The equation has been resolved for several isotopic species relevant for this work.
In contrast to the force-field approximation, widely used in CR astrophysics, the numerical approach presented here
is suitable to test arbitrary forms of the CR diffusion coefficient $K$, or to study mass/charge-dependent effects
arising from its functional dependence on rigidity or velocity.
In this work, we adopt a benchmark diffusion coefficient of the form
$K=K_{0}\beta(\R) \left(\R/{\R_{0}}\right)$
where, as we will discuss, the mass/charge dependence of $K(\R)$ are contained in the $\beta(\R)$ factor. 
The overall normalization is set by the $K_{0}$ parameter (at the reference rigidity $\R_{0}\equiv$\,1\,GV)
and it is of the order of $\sim{10^{22}}$\,cm$^{2}$s$^{-1}$.
In the follow, the main focus is given to the fluxes of total hydrogen
($^{1}$\Hyd+$^{2}$\Hyd) and total helium ($^{3}$\He+$^{4}$\He).

\section{Results and discussion}  
\label{Sec::Results}              

\subsection{Fitting the cosmic ray proton fluxes} 
\label{Sec::ProtonFits}                           
%
Our fits to the proton fluxes are shown in if Fig.\,\ref{Fig::ccEnergySpectra}.
In the 16 panels figure, the \AMS measurements of the proton fluxes are shown as open squares
for 16  BR's out of 79 \citep{Aguilar2018PHe}. 
The light orange lines describe our calculations for the best-fit proton fluxes, $\hat{J}_{\rm p}$.
The fits have been carried out independently for each BR using the least squares method.
For a given BR, the minimizing function is defined as:
\begin{equation}\label{Eq::Chi2}
\chi^{2}(k_{0}) = \sum_{j} \sigma_{j}^{-2} \left[   \hat{J}^{j}_{\rm p} - \frac{1}{\Delta\R_{j}}\int_{\R_{j}}^{\R_{j+1}} J_{\rm p}(\R)d\R \right]^{2}
\end{equation}
In the equation, a rigidity integration of the model output is computed within each bin $\Delta\R_{j}$, while for
the data the mean value is used. The $\sigma_{j}$ factors account for experimental uncertainties in the measured fluxes.
The theoretical flux is $J_{p}(\R)$, resulting from the Galactic LIS calculations of
Sect.\,\ref{Sec::Propagation} and from solar modulation model of Sect.\,\ref{Sec::SolarModulation}.
%
\begin{figure*}[!ht]
\centering
  \includegraphics[width=0.90\textwidth]{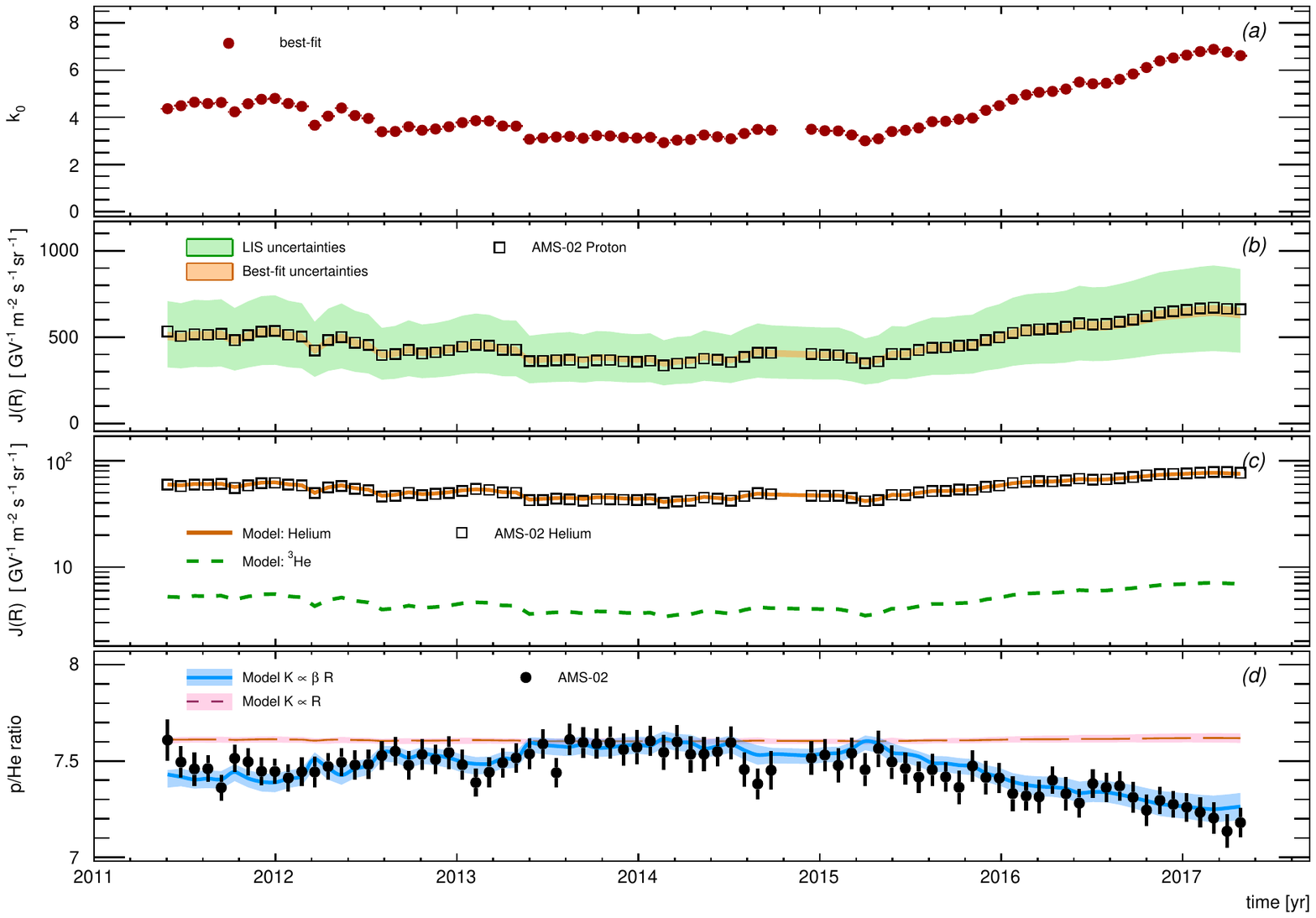}
\caption{ 
  (a) Time-series of the best-fit $k_{0}$-values derived with the \AMS proton
  data at $\R=1-60$\,GV \citep{Aguilar2018PHe}; 
  (b) time profile of the proton flux measured by \AMS 
  at $\R=$\,2\,GV in comparison with calculations, LIS uncertainties 
  and uncertainties from the fit;
  (c) time profile of the helium flux measured by \AMS
  at $\R=$\,2\,GV in comparison with best-fit calculations from the proton-driven fits (for $K=\beta\R$);
  (d) measured time profile of the \pHe{} ratio at $\R=$\,2\,GV in comparison
  with best-fit calculations for $K=\beta\R$ (thick solid line)
  and $K\propto\R$ (thin dashed line).
}
\label{Fig::ccPHeVSTimeBestFits}
\end{figure*}
%
In the solar modulation calculations, we have adopted a benchmark diffusion coefficient $K$ of the form
$K=K_{0}\beta \left(\R/{\rm{GV}}\right)$, where we express its normalization as
$K_{0}\equiv{10^{22}}k_{0}$\,cm$^{2}$s$^{-1}$, so that the adimensional factor $k_{0}$ is determined from the data. 
Because the fits are repeated for each BR, the procedure returns a time-series of best-fit values $\hat{k}_{0}$
describing the temporal evolution of CR diffusion coefficient. Such an evolution is at the basis of
the time-dependent nature of CR modulation \citep{Manuel2014,Bobik2016},
At this point, we note that $k_{0}$ is degenerated with the wind speed $V$. In particular,
the data are only sensitive to the combination $k_{0}/V$, so that
the use of different (or variable) $V$-values can be compensated by a proper rescaling of $k_{0}$.
Hence, the wind speed $V$ was kept fixed to the reference value of 400\,km\,s$^{-1}$.
In our model, the problem is modeled under the usual quasi-steady approach, \ie, by providing a time series
of steady-state solutions for $J_{\rm p}$ corresponding to a time series of input parameters $k_{0}$.
Such an approach is justified by the different timescales between the CR transport in the heliosphere and the changing solar activity.
In principle, the CR modulation phenomena cover large spectrum of timescales, from hours to years,
in relation to the physics process involved. In particular, the CR diffusion timescale decreases rapidly with increasing rigidity \citep{Strauss2011}.
At the typical rigidity of the \AMS data ($\R\gtrsim\,2$\,GV), the characteristic timescale of CR modulation does not exceed the monthly resolution of the data.
Moreover, the analyzed long-term \pHe{} behavior is manifested at yearly  timescales.
Nonetheless, we stress that each element of the best-fit time series $k_{0}(t)$ should be regarded as an effective scale value for
the diffusion coefficient (\ie, averaged over the CR propagation histories), not necessarily representing the instantaneous conditions of
the heliospheric turbulence \citep{Strauss2011}.
To test the robustness of the fitting results,
we also made specific checks for the possible influence of drift effects. 
For this purpose, we made dedicated runs within a formulation where the global effect of drift is captured by
an additional free parameter $v_{D}$ in the convective term, representing the
drift speed, without giving adiabatic cooling \citep{Jokipii1977,IsenbergJokipii1978}.
This causes no appreciable effects in the phenomenology of the \pHe{} ratio while, for the absolute fluxes
in the $\R\gtrsim$\,2\,GV region, the $v_{D}$ parameter is highly degenerated with the diffusion parameters.
Degeneracy between drift and diffusion is quite unavoidable in a 1D formulations,
and can be removed only in more spatially complex models.
Finally, to test the presence of undetected systematic bias in the physical inputs (or in the data),
we repeated the fitting procedure after leaving a global normalization factor as nuisance parameter.
%
\begin{figure*}[!ht]
\centering
  \includegraphics[width=0.94\textwidth]{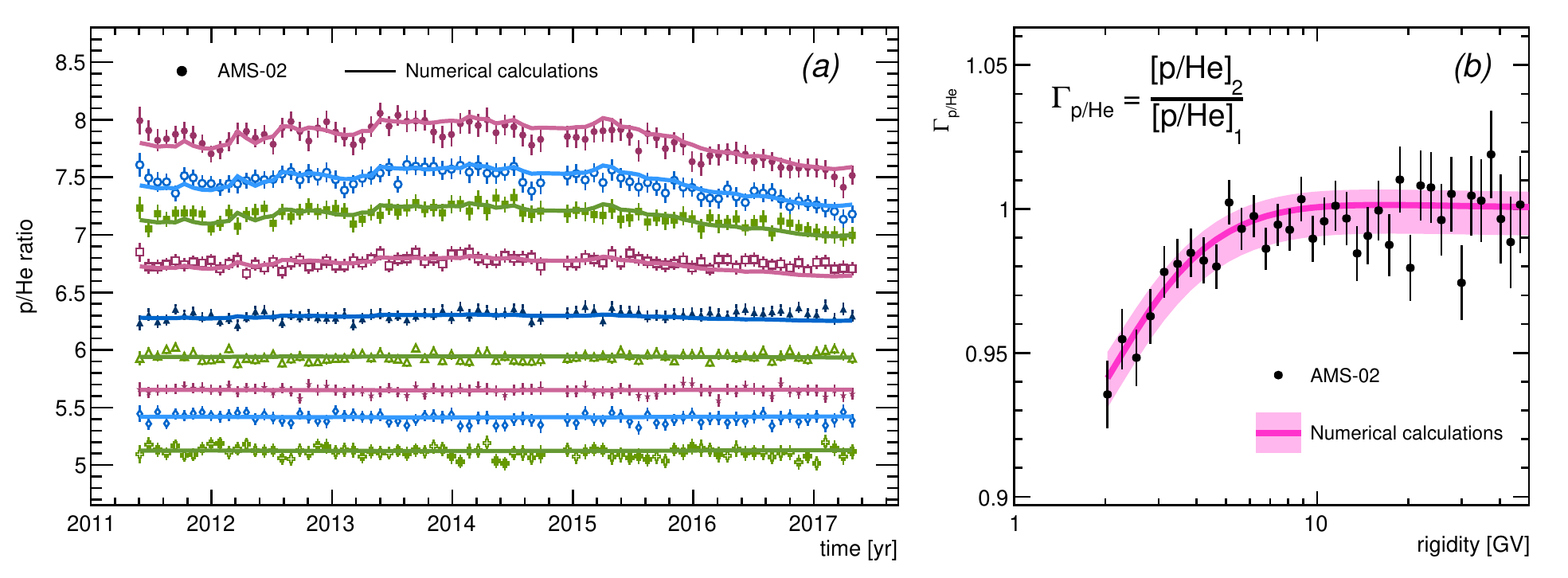}
\caption{
  (a) time profiles of the \pHe{} ratio evaluated the at rigidity $\R=$\,2.2, 2.5, 2.8, 3.4, 5.1, 7.4, 10.5, 15, and 22 GV (from top to bottom); 
  (b) rigidity dependence of the ratio $\Gamma_{\rm p/He}= {[\pHe]_{t_{2}}}/{[\pHe]_{t_{1}}}$
  calculated for February 2014 ($t_{1}$) and May 2017 ($t_{2}$).
}
\label{Fig::ccPHeRatioNC}
\end{figure*}
%

The best-fit time-series $\hat{k}_{0}$ 
is shown in Fig.\,\ref{Fig::ccPHeVSTimeBestFits}(a), along with the time profile of the CR proton fluxes at $\R=2$\,GV,
in Fig.\,\ref{Fig::ccPHeVSTimeBestFits}(b), in comparison with the \AMS data.
In the figure, the green shaded band represents the level of uncertainty on the proton LIS,
if it is directly propagated to the modulated spectra.
However, since $k_{0}$ is a free parameter, global variations in the proton LIS are partially re-absorbed
in the fit, \ie, by a proper scaling of the $k_{0}$-values and the nuisance parameters.
The relevant uncertainties in the proton flux are shown by the orange shaded band.
This band accounts for the standard uncertainties from the fitting method,
\ie, corresponding to one-sigma variations of $\chi^{2}$ curve as function of the free/nuisance parameters,
where the procedure is repeated for a large number of LIS functions.
In practice, the LIS-functions are Monte-Carlo generated according to the probability density functions of the underlying
key parameters (see Sect.\,\ref{Sec::Propagation}),
\ie, all input LIS's that produce the band of Fig.\,\ref{Fig::ccLISProtonHelium} \citep{Feng2016}.
With this procedure, uncertainties associated with the LIS modeling are included in
the final error band, with the proper account for their bin-to-bin correlation.
The orange band can also be regarded as the level of precision to which our LIS knowledge could
be reached, potentially, if the solar modulation is well understood and precisely modeled.

\subsection{Predicting the helium flux and the p/He ratio} 
\label{Sec::PHePrediction}                                 
%
Once the $k_{0}$ time-series is determined from the \AMS proton data,
we have used our proton-calibrated model to \emph{predict} the flux of CR helium.
This includes the calculation of $^{3}$\He{} and $^{4}$\He{} isotopes.
Helium flux calculations are shown in Fig.\,\ref{Fig::ccEnergySpectra} for 16 selected epochs.
The \AMS data on CR helium are shown as open circles. They are well matched by the
blue line,  showing the model prediction for the total helium flux (\ie, $^{3}$\He plus $^{4}$\He). 
The green band shows the sub-dominant $^{3}$\He{} component and its uncertainty.
Because $^{3}$\He{} is purely \emph{secondary} in CRs (\ie, it is produced by fragmentation of 
primary CR particles such as $^{4}$\He{} or $^{12}$\C), its calculations relies strongly on the
CR propagation parameters which, in turn, are constrained by the \BC{} ratio.
Thus, an improvement on the LIS helium calculation is due to the new \BC{} data from \AMS \citep{Aguilar2018LiBeB}.
From the proton-driven calculations of the helium flux, we made a corresponding  prediction for the \pHe{} ratio
and in particular for its temporal dependence.
The temporal dependence of CR helium at  $\R\approx$\,2\,GV is shown in  Fig.\,\ref{Fig::ccPHeVSTimeBestFits}(c),
where our predictions for the total flux are compared with the \AMS{} data. The
$^{3}$\He{} isotopic flux and its temporal dependence are also shown in the panel, as green dashed line.
Similarly, Fig.\,\ref{Fig::ccPHeVSTimeBestFits}(d) reports predictions and data for the \pHe{} ratio.
Along with the benchmark model, in Fig.\,\ref{Fig::ccPHeVSTimeBestFits}(d) 
we also plot the predictions from a diffusion coefficient of the form  $K{\propto}k_{0}\R$
(pink dashed line), \ie, without any $\beta$-dependence.
In this case there are no $A/Z$ terms in $K(\R)$, which is a sole function of rigidity, 
and it can be seen that the \pHe{} ratio is remarkably constant.
Within this model, the diffusion of proton and helium is identical, at a given rigidity,
so that any difference in their modulation
is only caused by their different LIS shapes.
This shows that the different spectral shapes play a minor effect in the \pHe{} time dependence.
On the other hand, before the recent measurements from \AMS and PAMELA,
the same  spectral shape was assumed for the protons and helium LIS.
The impact of \p-\He{} LIS differences in the solar modular is discussed in \citet{Gieseler2017,Corti2019,Webber2015}.

The \pHe{} predictions from our benchmark diffusion coefficient ($K\propto\beta\R$) are shown in the blue solid line.
It can be seen that, from this scenario, the predicted ratio agrees very well the observations.
The relevant uncertainties of the model predictions are shown as shaded bands.
They arise primarily from uncertainties in the \p-\He{} LIS models (as discussed above) and from
uncertainties on the $^{3}$\He{} components.
These contributions are linked to the Galactic CR propagation parameters. 
A minor contribution to the uncertainty band arises from the statistical errors of the fitting procedure.
Within the estimated uncertainties, our predictions of the temporal evolution of the \pHe{} evolution agree well with the \AMS data.

\begin{figure*}[!ht]
\centering
  \includegraphics[width=0.94\textwidth]{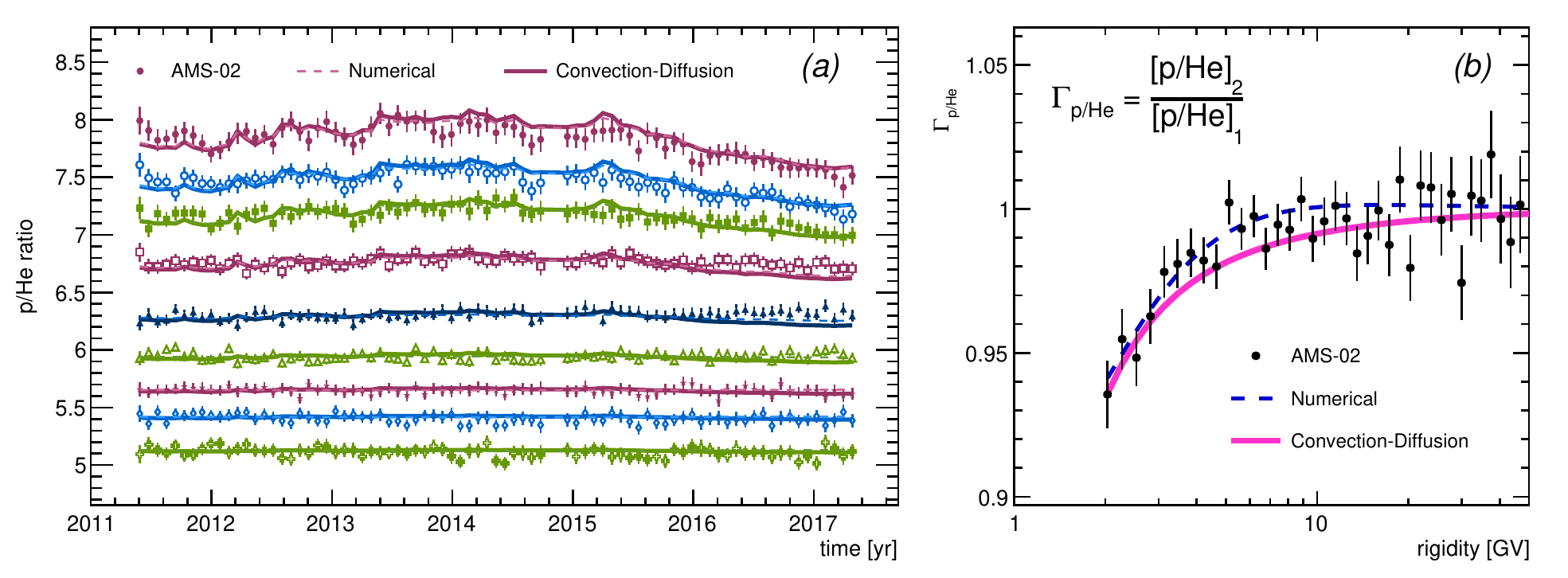}
\caption{
  (a) time profiles of the \pHe{} ratio evaluated at rigidity $\R=$\,2.2, 2.5, 2.8, 3.4, 5.1, 7.4, 10.5, 15, and 22 GV (from top to bottom); 
  (b) rigidity dependence of the ratio $\Gamma_{\rm p/He}= {[\pHe]_{t_{2}}}/{[\pHe]_{t_{1}}}$
  calculated for February 2014 ($t_{1}$) and May 2017 ($t_{2}$).
  In both plots, the new data from \AMS are compared with \emph{convection-diffusion} calculations.
}
\label{Fig::ccPHeRatioCD}
\end{figure*}

From the data, it can be noted that the \pHe{} behavior is correlated with the intensity of the individual \p-\He{} fluxes
Based on such a correlation, one may expect that the \pHe{} ratio is subjected to a cyclic modulation behavior, 
This is indeed the case, as shown in \citet{Boschini2019}. As results of simulations, the authors found that the \pHe{} ratio
is subjected to a quasi-periodical 11-year behavior. The absolute variation of the ratio is related to the level of solar activity.
With the present data, the variation in the \pHe{} ratio is observed  during the 
\emph{recovery phase} which begins approximately one year after the maximum of sunspots in Solar Cycle 24.
During this phase the CR fluxes increase to nearly a factor of $2$ in about two years.
The one year delay with respect to the 2014 solar maximum can be related to the CR modulation time lag. 
Such a lag can be interpreted as the time spent by the solar wind to transport the Sun's induced magnetic perturbations,
from the corona to the outer heliosphere \citep{Tomassetti2017TimeLag}.

\subsection{On the interpretation of the p/He behavior} 
\label{Sec::PHeInterpretation}                            

An important results of our work in \citep{Tomassetti2018PHeVSTime} is that the time evolution
of the low-rigidity \pHe{} ratio (as measured by \AMS at $\R\approx$\,3\,GV) is described very well with
a  $\beta\times\lambda(\R)$ dependence of the CR diffusion coefficient in the heliosphere.
This implies that the measured long-term dependence of the \pHe{} ratio is a signature of the
\emph{universal} rigidity dependence of the mean free path $\lambda(\R)$.
Given that the diffusion coefficient is $K(\R)=\beta\lambda(\R)/3$, 
the measured differences between the proton and helium flux evolution are caused by their
difference in the $A/Z$ ratio between charge and mass number. 
In the non-relativistic limit, proton  and helium nuclei at the same value of rigidity,
must travel at different speed with $\beta_{p}>\beta_{He}$. This is apparent from the expression of the CR speed as function of rigidity:
\begin{equation}\label{Eq::BetaVSRig}
 \beta(\R)= \R / \sqrt{ \R^{2} + (m_{p}A/Z)^{2}}
\end{equation}
Thus, at fixed rigidity, CR protons ($Z/A=$1) experience a slightly faster diffusion than helium ($Z/A\approx$1/2). 
On the other hand, in the relativistic limit one has $\beta\rightarrow$1, so that the diffusion of the two particles becomes identical.

The rigidity dependence is illustrated more clearly in Fig.\,\ref{Fig::ccPHeRatioNC}.
In the left panel, the \pHe{} time profile is plotted for several rigidity values from $\sim$\,2\,GV to $\sim$\,20\,GV
(from top to bottom). From this figure, the overall trend of the \AMS data is well described by the model.
It can also be seen that, for increasing rigidity, the \pHe{} ratio decreases rapidly from about $8$ to about $5$
(on average), following its interstellar rigidity dependence. 
The \pHe{} dependence upon rigidity is also shown in Fig.\,\ref{Fig::ccLISProtonHelium}(c).
Similarly, the temporal dependence of the \pHe{} ratio weakens as well at increasing rigidity,
so that at $R\gtrsim$\,3 GV the \pHe{} is essentially constant.
An illustration of the rigidity dependence is given in the right panel of Fig.\,\ref{Fig::ccPHeRatioNC}.
The plot shows the quantity $\Gamma_{\rm p/He}\equiv{[\pHe]_{t_{2}}}/{[\pHe]_{t_{1}}}$
computed at two epochs $t_{1}$ and $t_{2}$. 
As reference epochs we have chosen February 2014 ($t_{1}$, where the \pHe{} ratio is maximum) and May 2017 ($t_{2}$, where the ratio is minimum).
As shown by the solid line, the $\Gamma_{\rm p/He}$ function is predicted to increases monotonically with rigidity,
but it reaches a plateau as soon as the rigidity is relativistic (\ie, a few GV). 
This prediction agrees very well with the \AMS measurements.


\subsection{The convection-diffusion approximation} 
\label{Sec::CDA}                                    
%
\begin{figure*}[!ht]
\centering
  \includegraphics[width=0.94\textwidth]{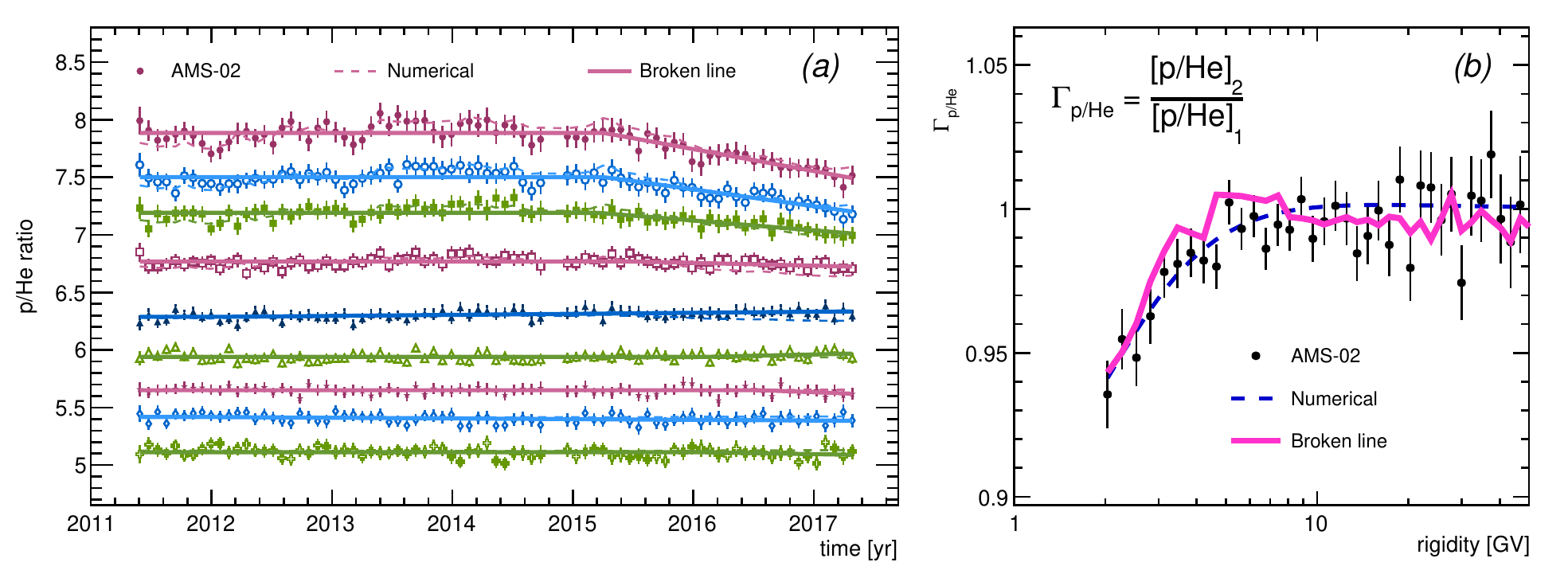}
\caption{
  (a) time profiles of the \pHe{} ratio evaluated the rigidity $\R=$\,2.2, 2.5, 2.8, 3.4, 5.1, 7.4, 10.5, 15, and 22 GV (from top to bottom); 
  (b) rigidity dependence of the ratio $\Gamma_{\rm p/He}= {[\pHe]_{t_{2}}}/{[\pHe]_{t_{1}}}$
  calculated for February 2014 ($t_{1}$) and May 2017 ($t_{2}$).
  In both plots, the new data from \AMS are compared with the \emph{broken line} fits.
}
\label{Fig::ccPHeRatioLF}
\end{figure*}
%

A simple and elegant interpretation of the \pHe{} behavior can be obtained within the \emph{convection-diffusion} approximation (CDA).
The CDA approach represents a low-order approximated solution of Eq.\,\ref{Eq::CRTransport},
where all spatial dependencies are neglected (0D).
In the CDA, the key hypothesis is that the
diffusive current of CR particles ($\sim{K}\partial\psi/\partial{r}$) is exactly countered by the convective current ($\sim{V}\psi$).
Any other process (such as magnetic drift or loss terms $\partial/\partial\rm{ln}\R$) are neglected.
Since the convective current moves always outward, following the solar wind, the key CDA hypothesis
implies that the diffusive current acts inward, which is not strictly true. 
From the simple balance of the two terms the transport equation becomes:
\begin{equation}
  \partial\psi/\partial{t}=  K\partial\psi/\partial{r} - V\psi
\end{equation}
In steady-state conditions, this equation has a simple analytical solution $\psi=\psi_{0}e^{-\M}$,
where $\psi_{0}$ is the LIS phase-space density for the considered CR particle type.
The changing conditions of the background plasma are captured by the adimensional parameter $\M$:
\begin{equation}
\M = \int_{r}^{d}\frac{V}{K}dr \approx \frac{Vd}{K} \approx \frac{Vd}{ K_{0} \beta\R/{\rm GV}}
\end{equation}
For typical values $V\sim$\,400\,km/s, $d\sim$\,100\,AU, and $K_{0}\sim$\,10$^{22}$\,cm$^{2}$/s,
the parameter $\M$ at the GV scale is of the order of unity. 
In practice, $\M$ is determined experimentally from the CR data for a given LIS hypothesis.
For a diffusion coefficient of the type $K\propto\,\beta\R$, it follows
that to first approximation the modulation strength is inversely proportional to $\beta\R$.
Within the CDA, the \pHe{} ratio is of the type:
\begin{equation}\label{Eq::pHeCDA}
  \pHe \approx \left(\frac{J_{\rm p}^{0}}{J_{\rm He}^{0}}\right) \frac{ e^{-Vd/K_{\rm p}(\R)} }{ e^{-Vd/K_{\rm He}(\R)} }
\end{equation}
where $K_{\rm p}$ and $K_{\rm He}$ are the diffusion coefficients of CR protons and helium nuclei, the difference of
which arises from the $A/Z$ factors contained in the CR velocity of Eq.\,\ref{Eq::BetaVSRig}.
The CDA solution shows explicitly that the modulation depends upon the parameter combination $\mu{\equiv}Vd/K_{0}$,
which captures the effects of changing conditions in solar activity and the underlying parameter degeneracy.
We have also verified that such a degeneracy is respect even by the numerical solution, 
as long as drift and latitudinal dependence are not considered.
In the limit $\mathcal{M}\lesssim$\,1, the \pHe{} ratio of Eq.\,\ref{Eq::pHeCDA} can be rewritten in the following form
\begin{equation}\label{Eq::CD}
\pHe \approx \frac{J_{\rm p}^{0}}{J_{\rm He}^{0}} \left\{ 1 -\frac{\mu(t)}{\R/{\rm GV}}\left[\frac{1}{\beta_{\rm p}(\R)}-\frac{1}{\beta_{\rm He}(\R)} \right] \right\}\,,
\end{equation}
that shows clearly the dependence of the ratio upon $\beta(\R)$. 
Similarly, the $\Gamma_{\pHe}(\R)$ function can be readily calculated as:
\begin{equation}\label{Eq::GammaCDA}
\Gamma_{\pHe}(\R) \approx
\frac{
1-\frac{\mu(t_{2})}{\R/{\rm GV}}\left[\frac{1}{\beta_{\rm p}(\R)}-\frac{1}{\beta_{\rm He}(\R)}\right]
}{
1-\frac{\mu(t_{1})}{\R/{\rm GV}}\left[\frac{1}{\beta_{\rm p}(\R)}-\frac{1}{\beta_{\rm He}(\R)}\right]
}
\end{equation}
From the above relations, one can easily understand the dependence of the \pHe{} ratio
upon the variable strength of CR modulation, expressed by the effective parameter $\mu(t)$,

Our calculations within the CDA framework are shown in Fig.\,\ref{Fig::ccPHeRatioCD}. Again, the
CR diffusion coefficient is of the type $K\propto{k_{0}}\beta\R$, where we used $A/Z\cong$\,1 for CR protons
and $A/Z\cong$\,3.8 for helium, to account for their average isotopic composition.
As discussed, the time dependence of the combination  $Vd/k_{0}$ is represented
by a time-series of $k_{0}$ parameters.
In the CDA calculations of Fig.\,\ref{Fig::ccPHeRatioCD}, the input time-profile $k_{0}=k_{0}(t)$
is taken from the fit results obtained in Sect.\,\ref{Sec::ProtonFits}. 
From the figure, the temporal evolution of the \pHe{} ratio is described fairly well within the CDA framework.
Similarly, a reasonable rigidity dependence for $\Gamma_{\pHe}(\R)$ is obtained.

From Eq.\,\ref{Eq::GammaCDA}, one may note that the $\Gamma_{\pHe}(\R)$ function appears as LIS-independent
within the CDA approach, due to the fact that the LIS part is factorized out in in Eq.\,\ref{Eq::pHeCDA}.
We emphasize, however, that the LIS dependence is implicit from the data-driven (\eg, proton driven)
determination of the parameter $\M$, and thus the CDA calculations for the \pHe{} evolution
\emph{do depend} on the assumed LIS.
On the other hand, the different LIS shapes for protons and helium
has an impact in the \pHe{} time evolution that cannot be accounted within the CDA framework.
Since energy losses are neglected, the CDA is also known to  systematically underestimate the low-energy CR fluxes in the inner heliosphere.
It follows that this approximation is far inadequate for describing the data at the accuracy level of \AMS.
The limitations of the CDA approximation are reviewed in details in Refs.\,\citep{Moraal2013,GleesonUrch1971}.

\subsection{The parametric description} 
\label{Sec::BrokenLine}                 

A parametric description for the evolution of the \pHe{} ratio can be obtained
by means of the simple \emph{broken-line} model  \citep{Aguilar2018PHe}:
\begin{equation}\label{Eq::LF}
  r(t,\R) = 
  \begin{cases}
    A(\R) & (t<t_{0}(\R))\\
    A(\R) + B(\R) \times (t-t_{0}(\R))  & (t\geq t_{0}(\R))
  \end{cases}
\end{equation}
In this function, $A=A(\R)$ is the average \pHe{} ratio value from May 2011 to $t_{0}$, 
being $t_{0}=t_{0}(\R)$ the time when the \pHe{} ratio deviates from the average $A(\R)$-value,
and $B(\R)$ is the slope of its temporal variation. 
Using Eq.\,\ref{Eq::LF}, we made fits to the \AMS data for every bin of rigidity between 2 and 60 GV.
At rigidity above 3.29\,GV, the ratio becomes consistent with a constant value $A$ at the 95\,\% confidence level,
\ie, it becomes time independent.
Below 3.29 GV, the observed ratio becomes steadily decreasing ($B<$\,0) after some epoch $t = t_{0}$.
The combined best-fit value for $t_{0}$ is February\,28\,$\pm$\,42\,days, 2015, at all rigidity values. 
This epoch coincides with the beginning of the \emph{recovery phase} of the CR fluxes.

The best-fit of the \pHe{} data using Eq.\,\ref{Eq::LF} is shown in Fig.\,\ref{Fig::ccPHeRatioLF}.
In the left panel, the best-fit \emph{broken-line} description of the \pHe{} behavior is shown in comparison with the data.
Our numerical calculations are also shown for comparison.
Overall, in spite of its lack of physical motivations, Eq.\,\ref{Eq::LF} provides a satisfactory parametric
description of the data at a fixed rigidity. In fact, the broken-line description was aimed at capturing
the gross feature of the \pHe{} evolution, \ie, its decreasing behavior during the recovery phase.
On the other hand, since the fits are made independently for each rigidity bin, the $\Gamma_{\pHe}(\R)$ behavior
shown in the right panel appears dominated by fluctuations of the same order of the data. 

In this work, we have provided a simple explanation for such a feature.
According to our calculations, the expected feature for the \pHe{} ratio is a progressive variation,
the amplitude of which depends on the changes in the CR diffusion coefficient.
In the period between March 2015 and May 2017, the variation is appreciable in the \AMS data.  
This period is the so-called ``recovery phase'' of the CR flux modulation,
\ie, when solar activity decreases rapidly toward the end of Solar Cycle 24.

\section{Conclusions}      
\label{Sec::Conclusions}   
%
In this paper we discussed in details the solar modulation calculations that we have
developed for interpreting the recent CR data released from \AMS.
To describe the data at precision level demanded by \AMS, we carried out improved calculations for the
interstellar fluxes of Galactic CR nuclei, along with their uncertainties,
and numerical calculations of CR propagation in the heliosphere
tightly calibrated against the new \AMS proton data.
Our data-driven approach enabled us to \emph{predict} the temporal dependence of the \pHe{} ratio
and to test specific forms for the diffusion coefficient of CRs in the heliosphere.
In the calculations discussed here there are numerous simplifications. For instance, we have neglected
the tensor nature of CR diffusion which is essential for describing the effects of anisotropic diffusion
or curvature drift. We have also introduced simplified description of the heliosphere, by neglecting
the latitudinal dependence of the solar wind, the termination shock, the heliospheric current sheet.
Accounting for all these features would inevitably increase the model complexity, thus requiring more parameters.
These simplification should be kept on mind in the interpretation of the basic parameters of the model,
as they should be regarded as effective values. For example, the neglection of magnetic drift may
affect the best-fit value for the diffusion parameter $k_{0}$. We expect that any drift-induced bias,
however, would be the same for CR proton and helium particles.
As we have discussed, the monthly-resolved measurements of the \pHe{} ratio are well suited to test CR diffusion in the heliosphere.
Being a ratio of positively charged particles, it cancels out potential biases arising from, \eg,
magnetic drift or anisotropic diffusion.

As we have show throughout this paper, the temporal behavior of the \pHe{} ratio is caused by the $A/Z$-dependence in the CR diffusion coefficient.
When expressed as function of rigidity, it can be written as $K(\R)= \beta(\R)\times\lambda(\R)/3$, where the
mean free path $\lambda$ is a universal function of rigidity,
so that the $A/Z$-terms are contained in the equation for the CR speed as function of rigidity $\beta(\R)$, Eq.\,\ref{Eq::BetaVSRig}.
The remarkable agreement between our prediction and the \AMS{} data supports the concept of \emph{universality} in the CR propagation histories in the heliosphere.
In the future, further tests will be made using data over a larger fraction of the Solar Cycle, or measurements of different nuclear ratios.

 \section*{Acknowledgements} 
%
B.B., E.F., and N.T. acknowledge support of \emph{Italian Space Agency} (ASI) under agreement \emph{ASI-UniPG 2019-2-HH.0}.
F.B. and M.O. acknowledge support of \emph{Portuguese Science Foundation} (FCT) under agreement \emph{CERN/FIS-PAR/0020/2017}.
The data used in this work were obtained from the
SSDC/ASI Cosmic-Ray Database
(\href{https://tools.ssdc.asi.it/CosmicRays}{\url{https://tools.ssdc.asi.it/CosmicRays}}).
\\



\end{document}